\documentclass{cimento} \input{psfig.sty} \title{Wind circumburst
density profile: a linear $E_{\rm p}$-$E_\gamma$ correlation}
\author{L.~Nava\from{ins:oab}\ETC, G.~Ghisellini\from{ins:oab},
  G.~Ghirlanda\from{ins:oab}, F.~Tavecchio\from{ins:oab} \atque
  C.~Firmani\from{ins:mx}}

\instlist{\inst{ins:oab} Osservatorio Astronomico di Brera, Via Bianchi
  46, I-23807 Merate Italy
  \inst{ins:mx} Instituto de Astronom\'ia, U.N.A.M., A.P. 70-264,
  04510, M\'exico, D.F., M\'exico}

\PACSes{\PACSit{98.70.Rz}{Gamma-Ray bursts}}

\begin{document}

\maketitle

\begin{abstract}
  Ghirlanda et al. (2004) derived the collimation-corrected energy
  $E_\gamma$ for a sample of 15 bursts under the assumption of a
  homogeneous circumburst density profile. They found a correlation
  (the so-called Ghirlanda correlation) between $E_\gamma$ and the
  rest frame peak energy of the $\nu F_\nu$ prompt spectrum ($E_{\rm
    p}$). Nava et al. (2006) showed that, assuming a circumburst
  density distribuited with a $r^{-2}$ wind profile, the Ghirlanda
  correlation remains tight and becomes linear. This implies that: i)
  it remains linear also in the comoving frame, no matter the
  distribution of bulk Lorentz factors, ii) it entails that different
  bursts have the same number of relevant photons. We have updated
  these findings including recently detected bursts (21 in total),
  stressing the two important implications.
\end{abstract}

\section{The $E_{\rm p}$-$E_\gamma$ correlation in the case of a homogeneous
  density profile} 
\begin{figure}[t!]
\hskip -0.4 cm
\psfig{figure=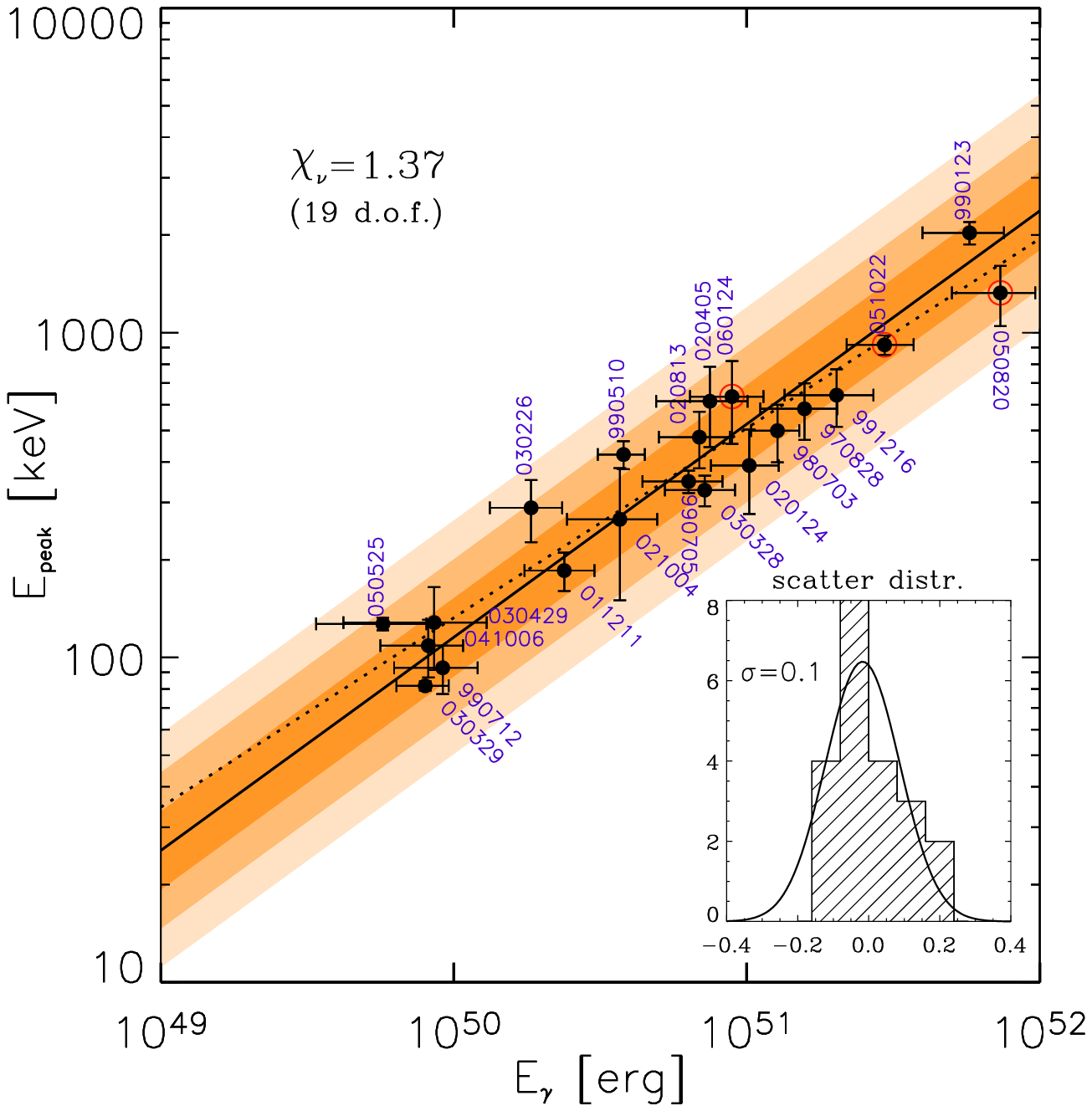,width=7cm,height=7cm} 
\hskip -0.2 cm
\psfig{figure=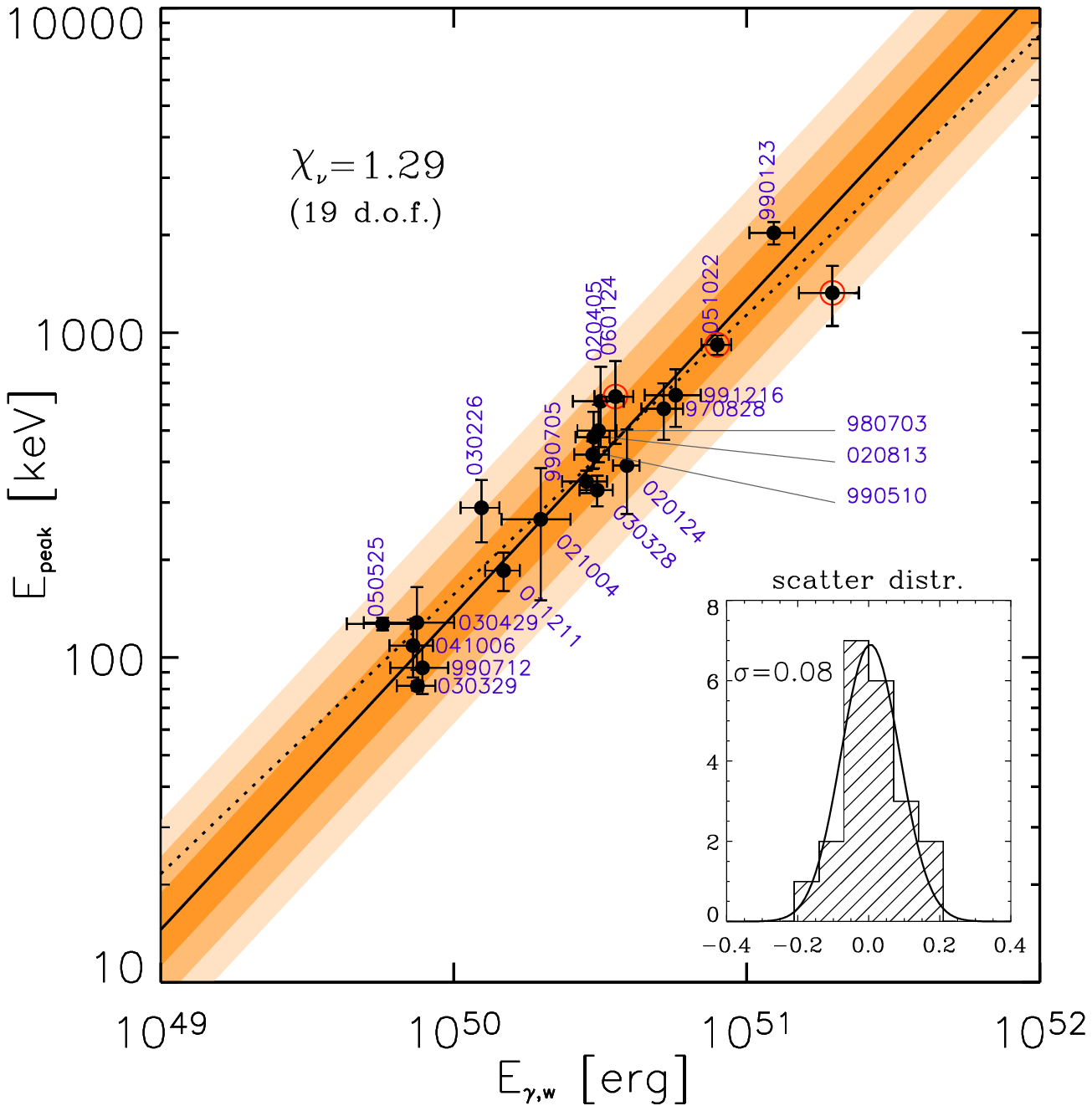,width=7cm,height=7cm}
\vskip -0.3cm
\caption{The updated Ghirlanda correlation between the rest frame peak
  energy and the collimation-corrected energy in the homogeneous case
  (left) and in the wind case (right). The data are listed in Tab. 1
  and Tab. 2 of \cite{ref:nava}. In addition there are three new
  bursts (circled points): GRB 050820a, GRB 051022 and GRB 060124.}
\label{fig}
\end{figure}

Under the assumption that GRBs are collimated
sources it is possible to estimate the semiaperture angle of the jet
$\theta_{\rm j}$. The main observational evidence in support of this
scenario is the presence of a break in the afterglow lightcurves. In
order to derive $\theta_{\rm j}$ we need to know the jet break time
$t_{\rm b}$, the circumburst density $n$ and the kinetic energy
left-over after the prompt phase. The relation linking these
quantities is different according to the circumburst density profile
(\cite{ref:sari},\cite{ref:cl}). Ghirlanda et al. (2004) assumed a
homogeneous density profile ($n= const$) and derived the jets opening
angles (from the measurements of the steepening in the optical
lightcurves for a sample of 15 GRBs (\cite{ref:ghirla}). They found
that $E_\gamma=E_{\rm \gamma,iso}(1-\cos\theta_{\rm j})$ correlate
with the peak energy $E_{\rm p}$ and this correlation is tighter than
the correlation between the isotropic energy $E_{\rm \gamma,iso}$ and
$E_{\rm p}$ (Amati correlation, \cite{ref:amati}). The updated version
of the Ghirlanda correlation is reported in Fig. \ref{fig}.

\section{The $E_{\rm p}$-$E_\gamma$ correlation in the case of a wind
  density profile} The assumption of a homogeneous circumburst medium
contrasts with the present model of long GRBs progenitors, which are
thought to be very massive stars. In fact, the intense stellar wind,
typical of very massive stars, should produce a circumburst density
profile which scales as $n\propto r^{-2}$. Nava et al. (2006) derived
the jet opening angle $\theta_{\rm j,w}$ (and therefore the
collimation-corrected energy $E_{\rm \gamma,w}$) for a sample of 18
bursts under the assumpion of a wind density profile. They
demonstrated that the collimation-corrected energy $E_{\rm \gamma,w}$
still correlates with $E_{\rm p}$ and this newly found correlation has
a slightly smaller scatter (see Fig. \ref{fig}) than that of the same
correlation found under the homogeneous medium assumption. The most
important property of the $E_{\rm p}-E_{\rm \gamma,w}$ relation in the
wind case is its linearity: $E_{\rm p}\propto E_{\rm \gamma,w}$.

The finding of a linear $E_{\rm p}$-$E_{\rm \gamma,w}$ correlation
reveals an intriguing property: since both $E_{\rm p}$ and $E_{\rm
  \gamma,w}$, being ``energies'', transform in the same way (i.e.
$\propto \Gamma^{-1}$) from the source rest frame to the comoving
frame, the slope of the correlation is ''invariant''. Moreover, the
linear $E_{\rm p} - E_{\rm \gamma,w}$ correlation implies that the
$E_{\rm \gamma,w}/E_{\rm p}$ ratio, which represents the number of
relevant photons, is constant from burst to burst and its value is
$\sim 10^{57}$. Finally, the tighter and steeper $E_{\rm p}-E_{\rm
  \gamma,w}$ correlation found in the wind case can be used to
constrain the cosmological parameters (see \cite{ref:cosmo2}),
similarly to what has done with the $E_{\rm p}-E_\gamma$ correlation
in the homogeneous case (\cite{ref:cosmo1}).

\end{document}